%% file: output.tex
\documentclass[preprint,12pt,authoryear]{elsarticle}

\usepackage{amssymb}
\usepackage{booktabs}
\usepackage{adjustbox}

\begin{document}

\begin{frontmatter}



\title{Gold Standard Pairs Trading Rules: Are They Valid?}


\author{Miroslav Fil}

\address{University of Oxford, Department of Computer Science}
\ead{miroslav.fil@seh.ox.ac.uk}

\begin{abstract}
Pairs trading is a strategy based on exploiting mean reversion in prices of securities. It has been shown to generate significant excess returns, but its profitability has dropped significantly in recent periods. We employ the most common distance and cointegration methods on US equities from 1990 to 2020 including the Covid-19 crisis. The strategy overall fails to outperform the market benchmark even with hyperparameter tuning, but it performs very strongly during bear markets. Furthermore, we demonstrate that market factors have a strong relationship with the optimal parametrization for the strategy, and adjustments are appropriate for modern market conditions.

\end{abstract}




\end{frontmatter}


\section{Introduction}

We examine the distance and cointegration methods in pairs trading, which is a mean-reversion contrarian trading strategy. It has been shown to perform well in historical backtests, achieving up to 11\% excess annualized return for US stocks over 1962-1997 \citep{gatev_pairs_1999}. Its performance has been significantly decreasing over time to the point where reasonable transaction costs often almost eliminate excess returns in more recent work \citep{do_does_2010, rad_profitability_2015}.

Nonetheless, the pairs trading strategy has been successful in a large number of national markets and timeframes \citep{jacobs_determinants_2015} despite its simplicity, performing best either in emerging markets or markets with a large number of potential pairs. The source and determinants of pairs trading profitability are not entirely clear, although the strategy has been noted to be sensitive to parametrization and transaction costs in particular \citep{huck_high_2013, do_are_2012}.

Pairs trading is a general framework and a great number of sophisticated approaches has been researched, including modeling the mean-reversion as an Ornstein-Uhlenbeck process \citep{elliott_pairs_2005}, stochastic control \citep{jurek_dynamic_2007}, genetic algorithms \citep{huang_intelligent_2015} and copulas \citep{liew_pairs_2013}. An exhaustive survey of pairs trading methods was recently developed by \citet{krauss_statistical_2017}. However, explicit comparisons between methods are scarce outside of the few most popular ones, and they do not seem to have a clear winner even when performed \citep{rad_profitability_2015, carrasco_blazquez_pairs_2018}.

This paper contributes in multiple areas. First, we conduct rigorous examination of the most common pairs trading methods with new trading data up to 2020 including the Covid-19 crisis. Second, we investigate the validity of several parameter choices ubiquitous across literature and demonstrate a relationship between transaction costs, execution lag and optimal parameters. Our research can thus serve as a guide for future work on pairs trading.

\section{Methodology}

\subsection{Data}
\label{section:data}
We use daily data on US stocks traded on NYSE spanning from 1990/01/01 to 2020/06/01. We further divide the history into sub-periods roughly demarcated by the dot-com and 2008 crises. The subperiods allow us to study the trading performance in bull and bear markets separately. In total, the dataset contains roughly 2800 stocks. We further pre-process the dataset by excluding bottom decile stocks in terms of liquidity, and removing any stocks that were not traded during the formation period for at least one trading day. Subperiod details can be found in Table \ref{tab:subperiods}.

\subsection{Methods}
Pairs trading proceeds in two stages, formation and trading. In the formation period, a statistical criterion is utilized to determine which pairs exhibit a stable long-run relationship. A measure of spread between the prices of securities making up the pair is defined, and its historical value is measured. In the trading period, positions are opened when the spread gets far enough from its estimated equilibrium. 

Pairs trading methods differ in how they form pairs, the trading rules used, or both. We apply the two most common pairs formation approaches, the distance and cointegration methods, which were first investigated by \citet{gatev_pairs_1999} and \citet{vidyamurthy_pairs_2004}, respectively.

The distance method proceeds by computing pairwise sums of squared deviations between the two normalized logarithmic price series, which we label as $P$ here:
\begin{equation}
    SSD_{ij}=\sum_t(P_{it}-P_{jt})^2
\end{equation}
\label{}
As suggested by \citet{gatev_pairs_1999}, we then pick the top 20 pairs minimizing this metric, indicating that their prices have evolved the most similarly. We define the spread as
\begin{equation}
    spread_{ijt}=P_{it}-P_{jt}
\end{equation}

In the trading period, we execute trades based on the spread's value normalized by the formation period. In particular, we short the spread when its normalized value exceeds 2, and close the position once it crosses zero or at the end of trading. We proceed analogically when it goes below -2. Such thresholded trading rules are very common in pairs trading literature and they directly replicate the setup from \citet{gatev_pairs_1999}.

The cointegration method is similar conceptually. The Engle-Granger cointegration test \citep{engle_co-integration_1987} is used to filter out cointegrated pairs. The spread is defined to be the residual from the Engle-Granger cointegration regression. Trading then proceeds exactly the same as in the distance method.

Again following \citet{gatev_pairs_1999}, we also utilize a one-period trade execution lag in order to estimate the bid-ask spread and other trade execution difficulties. We also employ a computationally efficient alternative to the cointegration method inspired by \citet{rad_profitability_2015}. We pre-determine the desired number of pairs, sort all pairs by the sum of squared deviations metric used in the distance method, and keep testing the most similar pairs for cointegration only until we achieve the desired number of pairs.

\subsection{Transaction costs}
\label{section:txcost}
Transaction costs for US stocks in the pairs trading context were thoroughly examined by \citet{do_are_2012} who split transaction costs among commissions, market impact and short-selling costs. They all vary over time, but they have generally been decreasing. We further extrapolate the provided analysis and work with one-way transaction costs ranging from 35bps in 1990 to 26bps in 2020. Short-selling costs follow the work of \citet{davolio_market_2002} who found an average fee of 0.6\% p.a. for US equities in 2000-2001, which we apply over the whole period.

A summary of transaction costs is available in Table \ref{tab:subperiods}.

\begin{table}[ht!]
    \small
    \begin{tabular}{lllllll}
    \toprule
    {} &   1990-00 &   2000-02 &   2002-07 &   2007-09 &   2009-20 & Covid \\
    \midrule
    Start date &  1990/1/1 &  2000/3/1 &  2002/10/1 &  2007/8/1 &  2009/6/1 &   2020/2/20 \\
    End date   &  2000/3/1 &  2002/10/1 &  2007/8/1 &  2009/6/1 &  2020/2/20 &   2020/6/1 \\
    Tx. cost      &      35bps &       30bps &       30bps &       30bps &      26bps &       26bps \\
    Short cost    &        0.6\% &        0.6\% &        0.6\% &        0.6\% &        0.6\% &         0.6\% \\
    \bottomrule
    \end{tabular}
    \caption{Overview of subperiods}
\label{tab:subperiods}
\end{table}

\section{Results}
In order to evaluate the viability of the pairs trading strategy, we conduct an extensive grid search across hyperparameters. We contrast them with the basic parameters originally used by \citet{gatev_pairs_1999} which are commonplace in pairs trading literature.

All the tried parameters are outlined below, and the most common standard values are highlighted in bold. All combinations of parameters were backtested. Of particular note is the period length multiplier, which is a scaling coefficient for the length of both the formation and trading periods. \citet{gatev_pairs_1999} originally chose 12-months formation and 6-months trading, which here has coefficient 1, and a coefficient of 0.5 means 6-months formation and 3-months trading, for example.

\begin{itemize}
    \item \# of nominated pairs - 5, 10, \textbf{20}, 40
    \item trading trigger threshold - 0.5, 1, 1.5, \textbf{2}, 2.5, 3
    \item period length multiplier - 0.16, 0.5, \textbf{1}, 1.5
    \item conf. level of statistical tests (coint. method only) - 0.01, \textbf{0.05}, 0.1
\end{itemize}

Table \ref{tab:base_scenarios} shows an overview of all the main results. Returns utilizing the baseline parameters are specified as "base", and we also provide the best achievable returns in each subperiod based on our parameter search marked as "best". "Adaptive" and "finetuned" are other series of returns described in Section \ref{section:adaptive}. The Total column refers to results computed over the whole 1990-2020 period.

\subsection{Base scenarios}
Results with the baseline parameter settings show that pairs trading generally fails to perform in both absolute and risk-adjusted terms when compared to the market benchmark. It does however perform remarkably well during bear markets, producing excess returns on the order of 2\% per month, although it lags close to 1\% behind the market in bull markets. This behavior is consistent across all the trading history from 1990 to 2020 including the Covid-19 crisis. The risk-adjusted returns follow the same pattern as the raw returns.

Overall, the consistent underperforming of the market during bull markets leads to the base strategy producing less than 0.2\% monthly returns over the whole period in comparison to the market's 0.47\%. As such, it appears that naive pairs trading is not viable unless we only utilize it during bear markets. However, our subperiod separation utilizes significant foresight bias in knowing the exact beginnings and ends of bear markets. In practice, it is not trivial to determine when a bear market is taking place, which makes well-timed application of pairs trading difficult.

\begin{table}
\footnotesize
\setlength{\tabcolsep}{4pt}
\input{tables/reshaped_results}
\label{tab:base_scenarios}
\caption{Returns across pairs trading methods and timeframes}
\end{table}

\subsection{Adaptive trading}
\label{section:adaptive}
In Table \ref{tab:base_scenarios}, we also show the maximum possible returns achievable (marked as "best") in the given subperiod. Those returns use parameters that can only be determined after the subperiod is over. We compare them with a strategy that adapts the strategy parameters every two years to make them optimal for the preceding two-year period, and then aggregate those results to conform to the subperiods introduced in Section \ref{section:data}. The latter strategy is actually executable because it requires no foresight bias, and is denoted as "adaptive". 

Nonetheless, the adaptive trading strategy performs worse than even the baseline parameters. However, we see from the "best" results that if we were able to trade better in the trading period, the returns would significantly improve, making the strategy perform well even during bull markets. The cointegration method in particular appears to have a higher ceiling for performance since it can achieve up to 1.03\% monthly profit over the whole history as compared to the distance method's 0.64\%, which is only slightly higher than the market's overall monthly return of 0.47\%. 

Table \ref{tab:params} shows the averaged optimal parameters of top 3 parameter settings for each period. The adaptive strategy essentially tries to exploit momentum in parameter values, but that does not seem to be feasible. We see that the parameters are fairly volatile over time, although there do appear to be significant differences from the baseline parameters set by \citet{gatev_pairs_1999}.

We note that the optimal parameters vary quite strongly across different subperiods, and the best parameters for the whole examination period are not necessarily close to the average of subperiod optimal parameters as one might expect. Additionally, the optimal parameters are not very stable even when considered only across each of bull and bear market subperiods, further indicating the difficulty of finding well-performing static trading rules.

Utilizing the optimal parameters for backtesting over the whole period from Table \ref{tab:params}, we construct another parametrization denoted as "finetuned" in Table \ref{tab:base_scenarios}. We still observe a significant gap between those and the maximum achievable returns when continually using the best parameters for each subperiod. Therefore, even with significant foresight bias and actually knowing which static parameters would be the best over the whole period, the returns still pale in comparison to dynamically chosen parameters. In fact, completely static trading rules of any sort are not enough to outperform the market benchmark since the distance and cointegration methods then only achieve 0.23\% and 0.26\% monthly return, respectively.

\begin{table}
\small
\setlength{\tabcolsep}{4pt}
\input{tables/best_param_avgs_separate}
\label{tab:params}
\caption{Optimal parameters across pairs trading methods and timeframes}
\end{table}

\subsection{Execution lag and transaction costs}
Table \ref{tab:lag_txcost} shows the optimal strategy parameters across a range of lags and transaction costs. Each table entry is the average across all history of the top 3 optimal parameter settings in each subperiod. 

We observe that as the transaction costs rise, it becomes advantageous to increase the threshold and period length. The number of pairs and confidence level are however much less affected. Furthermore, increasing the lag from zero to one sharply increases the optimal number of pairs, although the effect on other parameters is fairly weak.

Overall, the results here can be seen as motivating for parameter choices. \citet{gatev_pairs_1999} explicitly considered one-period execution lag and the transaction costs at the time of his analysis were mostly even higher than 30bps. As Table \ref{tab:lag_txcost} shows, the choice of 20 top pairs with 2 standard deviation threshold and period length multiplier of 1 are quite close to optimal for the distance method under those conditions. However, it also seems that as the transaction costs are getting lower in modern times and trade execution difficulties, for which the execution lag is a proxy, are becoming less severe, the standard pairs trading parameters need to be adjusted accordingly.

For example, given that market impact costs are a significant component of transaction costs as discussed in Section \ref{section:txcost} and further considering that market impact costs are likely to scale with employed capital, we can hypothesize that different parametrization would be ideal contingent on how much capital is available for trading. Since less capital implies lower market impact, which in turn means lower total transaction costs, we would expect lower thresholds and shorter trading/formation periods to perform better in those conditions as per our discussion above.

\begin{table}

\input{tables/merged_lag_txcost}
\label{tab:lag_txcost}
\caption{Dependence of optimal parameters on transaction costs and execution lag}
\end{table}

\section{Conclusion}
We confirm that pairs trading does not have high excess returns in recent times. We further corroborate the findings of \citet{do_does_2010} that pairs trading performs very strongly during financial crises including the Covid-19 crisis, but generally lags behind the market benchmark otherwise, both in raw and risk-adjusted returns. This leads to overall weak performance.

However, it would seem that naive pairs trading is not entirely defunct at least in the formation stage, as solid returns can be achieved with optimal trading rules even when using basic formation strategies. The viability of such simple pairs trading strategies is usually examined in the context of thresholded trading rules which tend to be directly replicated from previous work. However, we show that their optimal parametrization shows a strong relationship with market conditions, and can thus be adapted to the market at hand. While the baseline parameters were close to optimal in the market conditions present at the time of the original analysis by \citet{gatev_pairs_1999}, we present evidence that they are essentially outdated nowadays, especially for the cointegration method.

Our findings indicate that studies directly replicating the pairs trading setup from \citet{gatev_pairs_1999} are not perfectly representative of the actual quality of the strategy. The parameters therein were implicitly optimized for the market conditions at the time. They are not necessarily valid in recent periods, and their suboptimality is likely to play a role in the widely held belief that simple pairs trading no longer works. 

However, finding good static trading rules that would consistently and significantly outperform the gold standard trading rules appears hard to do, at least without resorting to sophisticated methods. In fact, it appears that it might not be possible unless we use at least partially dynamic trading rules. We thus conclude that naive pairs trading with complexity on the level of \citet{gatev_pairs_1999} does not work well, regardless of whether hyperparameter tuning is performed or not, and its weak performance cannot be fixed with better parameters alone since they do not bring large enough benefit. 

\section*{Acknowledgments}
This research did not receive any specific grant from funding agencies in the public, commercial, or not-for-profit sectors.



\bibliographystyle{elsarticle-harv}
\bibliography{pairs_trading_paper2}

\end{document}

%% file: tables/reshaped_results.tex
\begin{tabular}{lllllllll}
\toprule
       &                   & 1990-00 & 2000-02 & 2002-07 & 2007-09 & 2009-20 & Covid & Total \\
\midrule
Dist. & Mo. profit (base) &    0.16\% &    0.25\% &    0.07\% &    0.40\% &    0.05\% &       0.94\% &    0.11\% \\
       & Mo. profit (best) &    0.88\% &    0.90\% &    0.50\% &    1.18\% &    0.30\% &       1.30\% &    0.64\% \\
       & Mo. profit (adaptive) &   -0.03\% &    0.46\% &    0.10\% &    0.64\% &   -0.08\% &      -0.47\% &    0.05\% \\
       & Mo. profit (finetuned) &    0.38\% &    0.96\% &    0.10\% &    0.58\% &   -0.08\% &       0.27\% &    0.23\% \\
       & Ann. Sharpe (base) &     -0.07 &       0.3 &      -1.3 &         1 &      -1.6 &          2.2 &     -0.86 \\
       & Ann. Sharpe (best) &       1.4 &       1.9 &       1.4 &       1.2 &     -0.47 &          5.4 &      0.81 \\
       & Ann. Sharpe (adaptive) &     -0.68 &      0.78 &      -1.7 &      0.29 &        -2 &         -1.3 &     -1.16 \\
       & Ann. Sharpe (finetuned) &      0.29 &       1.5 &     -0.78 &     -0.09 &      -1.4 &         0.15 &      -0.4 \\
Coint. & Mo. profit (base) &    0.24\% &    0.59\% &    0.08\% &    0.41\% &    0.14\% &      -0.72\% &    0.18\% \\
       & Mo. profit (best) &    0.84\% &    2.04\% &    0.54\% &    2.06\% &    1.00\% &       1.99\% &    1.03\% \\
       & Mo. profit (adaptive) &   -0.05\% &    0.89\% &   -0.51\% &   -0.10\% &   -0.05\% &      -1.85\% &   -0.11\% \\
       & Mo. profit (finetuned) &    0.50\% &    1.45\% &   -0.09\% &    0.67\% &   -0.12\% &      -2.04\% &    0.26\% \\
       & Ann. Sharpe (base) &      0.48 &       1.8 &     -0.89 &      0.72 &     -0.64 &         -3.1 &     -0.27 \\
       & Ann. Sharpe (best) &       1.8 &       4.5 &       1.5 &       1.4 &      0.63 &          4.1 &      1.57 \\
       & Ann. Sharpe (adaptive) &     -0.47 &         2 &      -4.7 &      -2.1 &      -2.6 &         -1.4 &     -1.96 \\
       & Ann. Sharpe (finetuned) &       1.2 &       2.8 &      -1.7 &      0.97 &      -2.3 &         -2.6 &     -0.41 \\
Market & Mo. profit &    0.90\% &   -0.83\% &    1.15\% &   -1.95\% &    0.63\% &      -4.79\% &    0.47\% \\
       & Ann. Sharpe &      0.64 &      -1.7 &       1.1 &      -1.1 &      0.94 &        -0.86 &      0.38 \\
\bottomrule
\end{tabular}

%% file: tables/best_param_avgs_separate.tex
\begin{tabular}{llrrrrrrr}
\toprule
       &           &  1990-00 &  2000-02 &  2002-07 &  2007-09 &  2009-20 &  Covid &  Total \\
\midrule
Dist. & \# of pairs &      13.67 &      13.33 &       7.22 &      28.33 &      15.33 &          8.33 &      23.33 \\
       & Period mult. &       0.31 &       0.44 &       0.98 &       0.16 &       0.47 &          0.67 &       0.16 \\
       & Threshold &       1.53 &       0.83 &       1.22 &       0.67 &       1.57 &          2.17 &       0.50 \\
Coint. & \# of pairs &       7.33 &       5.00 &       6.67 &       5.00 &       7.00 &          5.00 &       6.67 \\
       & Period mult. &       0.37 &       0.50 &       0.92 &       0.16 &       0.61 &          0.16 &       0.50 \\
       & Confidence &       0.06 &       0.01 &       0.07 &       0.05 &       0.04 &          0.10 &       0.01 \\
       & Threshold &       2.03 &       1.50 &       1.44 &       2.50 &       1.80 &          2.83 &       2.67 \\
\bottomrule
\end{tabular}

%% file: tables/merged_lag_txcost.tex
\begin{tabular}{llllrrr}
\toprule
       &     &   &           & \multicolumn{3}{l}{Transaction costs} \\
       &     &   &           &            0.000 &  0.003 &  0.005 \\
\midrule
Dist. & Lag & 0 & \# of pairs &            11.67 &   5.83 &   8.33 \\
       &     &   & Period mult. &             0.33 &   0.80 &   1.14 \\
       &     &   & Confidence &             0.05 &   0.05 &   0.05 \\
       &     &   & Threshold &             0.67 &   1.25 &   1.75 \\
       &     & 1 & \# of pairs &            25.00 &  18.33 &  16.67 \\
       &     &   & Period mult. &             0.22 &   0.86 &   1.08 \\
       &     &   & Confidence &             0.05 &   0.05 &   0.05 \\
       &     &   & Threshold &             0.67 &   1.33 &   2.25 \\
Coint. & Lag & 0 & \# of pairs &             5.00 &   5.83 &   5.83 \\
       &     &   & Period mult. &             0.27 &   0.47 &   0.72 \\
       &     &   & Confidence &             0.08 &   0.06 &   0.05 \\
       &     &   & Threshold &             1.67 &   2.08 &   2.42 \\
       &     & 1 & \# of pairs &            11.67 &  10.00 &   7.50 \\
       &     &   & Period mult. &             0.27 &   0.69 &   0.69 \\
       &     &   & Confidence &             0.08 &   0.04 &   0.04 \\
       &     &   & Threshold &             1.42 &   2.25 &   2.58 \\
\bottomrule
\end{tabular}